\documentclass[aps,prl,twocolumn,showpacs,superscriptaddress]{revtex4}

\usepackage{graphicx} 
\usepackage{longtable}

\begin{document}

\title{Chiral single-wall  gold nanotubes}
\author{R. T. Senger}\email{senger@fen.bilkent.edu.tr}
 \affiliation{Department of Physics, Bilkent University, 06800 Ankara, Turkey}
 \affiliation{T{\"U}B{\.I}TAK - UEKAE, P.K.74, 41470 Gebze, Kocaeli, Turkey}
\author{S. Dag}\email{sefa@fen.bilkent.edu.tr}
 \affiliation{Department of Physics, Bilkent University, 06800 Ankara, Turkey}
\author{S. Ciraci}\email{ciraci@fen.bilkent.edu.tr}
 \affiliation{Department of Physics, Bilkent University, 06800 Ankara, Turkey}

\date{\today}

\begin{abstract}
Based on first-principles calculations we show that gold atoms can
form both free-standing and tip-suspended chiral single-wall
nanotubes composed of helical atomic strands. Free-standing,
infinite (5,5) tube is found to be energetically the most
favorable.  While energetically less favorable, the experimentally
observed (5,3) tube stretching between two tips corresponds to a
local minimum in the string tension. Similarly, the (4,3) tube is
predicted  as a favorable structure yet to be observed
experimentally.
 Analysis of band structure, charge density, and quantum ballistic
conductance suggests that the current on these wires is less
chiral than expected, and there is no direct correlation between
the numbers of conduction channels and helical strands.
\end{abstract}


\maketitle

Current trends in   miniaturization of electronic devices have
motivated a growing interest in various nanoscale structures. Of
these structures nanowires constitute an important class in
nanoelectronics with their potential applications as nanodevices
or as connectors between them. Properties of very thin metal wires
are actively studied both experimentally and
theoretically\cite{gimzewski,ciraci89,agrait93,pascual93,mehrez97,gulseren98,sorensen98,kondo00,tosatti01,oshima03};
their formation  in the form of coaxial shells having helical
structures, as well as single atomic chains hanging between two
electrodes have been predicted for Cu \cite{mehrez97}, Al and Pb
\cite{gulseren98}, and Au \cite{sorensen98}. Synthesis of single
Au atom chain suspended between two Au electrodes has been a real
breakthrough in nanotechnology \cite{ohnishi98,yanson98}.
Furthermore, in UHV-TEM experiments it has been shown that Au
nanobridges can transform into several nanometers long regular
chiral nanowires suspended between two Au electrodes.
Interestingly, in agreement with previous theoretical studies,
these thin nanowires have the form of helical multiwall structures
of specific ``magic" sizes \cite{kondo00}.

\begin{figure}
\includegraphics[scale=0.36]{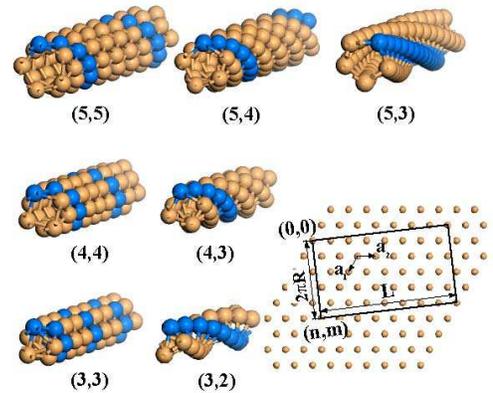}
\caption{Atomic structure of $(n,m)$ gold SWNTs obtained by
cylindrical folding of the 2D triangular lattice. Basis vectors of
the 2D lattice are ${\rm \bf a}_1$ and ${\rm \bf a}_2$; the chiral
vector  ${\rm \bf C}=n{\bf a}_1 + m{\bf a}_2$, such that the tube
circumference is $|{\bf C}|$, and radius
$R=(n^{2}+m^{2}-nm)^{1/2}|\bf{ a}_1|/2\pi$. An $(n,m)$ tube has
$n$ helical strands and $m$ defines the chirality.  A  dark
helical strand of atoms highlights the chirality of the tubes. The
radius $R$ and period $L$ are modified upon relaxation of the
tubular structure.} \label{fig:str}
\end{figure}

In recent UHV-TEM experiments evidence is found for the formation
of Pt and Au  single-wall nanotubes (SWNT): For Pt, the tubes
consist of 5 or 6 atomic rows that coil helically around the axis
of the tube \cite{oshima02}. In the case of gold,
 the SWNT was observed to be   composed of 5 helical
strands \cite{oshima03}. The shell of those SWNTs can be
constructed from rolling of a triangular network of gold atoms
onto a cylinder of radius $R$ as described in Fig.~\ref{fig:str}.
 Similar to carbon nanotubes,
the $(n,m)$ notation defines the structure of the tube.  According
to the work by Oshima {\it et al.}\cite{oshima03} the $(5,3)$ SWNT
(without a linear strand at the center) was  a long-lived
metastable structure that has been observed between $(7,3)$ wire
(with a strand at the center) and single Au atomic chain
synthesized during electron beam thinning of Au thin foil. Apart
from being the first observation of a Au chiral SWNT, this
interesting result has posed several important questions  as to
why only $(5,3)$ tube is observed among tubes with $n=5$; what
other tubes with $n<5$ are metastable. While the $(5,3)$ Au tube
being attached to electrodes is metastable, can free-standing Au
SWNTs be stable in the absence of central linear strand? How does
the chirality of helical strands influence the ballistic
transport? So far neither these questions have been addressed, nor
the existence of gold SWNTs has been theoretically demonstrated.

In this letter, we show that tubular structures of gold  can
indeed exist and display interesting electronic and transport
properties. Moreover, we explain why only a specific SWNT
suspended between two gold tips has been observed experimentally.
As single-wall tubular structures of gold, we considered all
possible cases with $n=5,4,3$ and $n\geq m > n/2$, since $(n,n-m)$
tubes are equivalent to $(n,m)$ with opposite chirality, and tubes
with $n\geq6$ have large enough radii to accommodate an extra
linear strand of gold atoms, therefore lacking a tubular
character. In particular, $(4,2)$ structure also has a non-tubular
form corresponding to a dumbbell chain and is not included in our
considerations. The structure of $(5,3)$ tube deduced
experimentally has a period approximately five times larger than
the lattice parameter of the ideal tube rolled from the undeformed
gold sheet \cite{oshima03}. The periodicity length $L$ of the
chiral tubes can be altered by a small axial shear. This is
achieved by applying a shear strain of $\epsilon
\parallel {\bf a}_2$ in the triangular network. We take $\epsilon
=0$ to achieve the periodicity of the $(5,3)$ tube through
$2\pi/5$ rotation of helical strands and hence to reduce the
number of atoms in the supercell from 190 to 38. Omitting such a
small strain ($\epsilon=0.005$ for $(5,3)$ tube \cite{oshima03})
does not affect our conclusions, but cuts down computational
effort dramatically. We carried out total energy and electronic
structure calculations on seven different tubular structures shown
in Fig.~\ref{fig:str}, using first-principles pseudopotential
plane-wave method \cite{vasp} within generalized gradient
approximation (GGA). All the atomic positions and lattice
parameters of tubular structures have been optimized through
lowering total energy, atomic force, and stress. The stability of
relaxed structures are tested also by \emph{ab-initio} molecular
dynamic calculations carried out at $T=800$ K. Spin-relaxed
calculations yielded zero total magnetic moments  for the
structures in Fig.~\ref{fig:str}. The analysis of quantum
ballistic conductance has been performed by using an
\emph{ab-initio} transport software based on localized basis sets
and non-equilibrium Green's function formalism \cite{trans}. The
structural properties of the optimized gold SWNTs are summarized
in Table~\ref{tab:res}.

\begin{table}
\caption{\label{tab:res} Structural  properties and energetics  of
relaxed chiral  gold nanotubes $(n,m)$. There are $N$ atoms in one
unit cell  which has length $L$ and radius $R$, both expressed in
\AA~units. $E_{b}$ and $E_c$ are the binding and curvature
energies per atom in units of eV, respectively. The string tension
of the tip-suspended tube, $f$, has units eV/\AA. }
\begin{tabular}{crrcccc} \hline \hline
   Structure   & ~~$N$ & ~~~~~$L$~~ & ~~~$R$~~~ & ~~~$E_b$~~~ & ~~~$E_c$~~~ & ~~~$f$~~~   \\ \hline
    (5,5)      & 10    &  4.63      & 2.44      & 2.66        & 0.18        &  1.188       \\
    (5,4)      & 14    &  7.15      & 2.24      & 2.60        & 0.24        &  1.179       \\
    (5,3)      & 38    & 20.73      & 2.12      & 2.58        & 0.26        &  1.154       \\
    (4,4)      &  8    &  4.60      & 2.04      & 2.54        & 0.30        &  1.166       \\
    (4,3)      & 26    & 16.91      & 1.85      & 2.52        & 0.32        &  1.062       \\
    (3,3)      &  6    &  4.39      & 1.71      & 2.41        & 0.43        &  1.083       \\
    (3,2)      & 14    & 12.28      & 1.51      & 2.31        & 0.53        &  1.017       \\ \hline \hline
\end{tabular}
\end{table}

\begin{figure}
\includegraphics[scale=0.38]{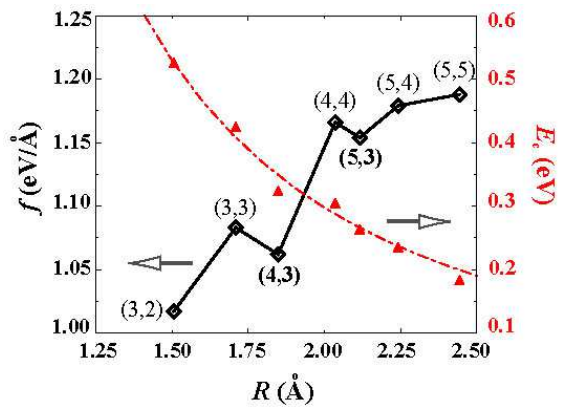}
\caption{ Triangles are calculated curvature energy $E_{c}$ of
free-standing SWNT. Dash-dotted curve is the best fit in the form
$\alpha/R^2$ ($\alpha=1.19$ eV \AA$^2$). Diamonds are  calculated
string tension $f$ of suspended $(n,m)$ gold nanotubes at zero
temperature. The local minima of $f$ indicate that $(5,3)$ and
$(4,3)$ SWNTs  are magic.} \label{fig:f}
\end{figure}

The binding (or cohesive) energy $E_{b}(n,m)= E_{T}(A)-
E_{T}(n,m)/N$, is calculated as the difference between the energy
of single Au atom, $E_{T}(A)$, and the total energy (per atom) of
the fully relaxed $(n,m)$ tubular form  having $N$ atoms in the
unit cell, $E_{T}(n,m)/N$. Accordingly, $E_{b}>0$ (exothermic)
indicates a stable structure corresponding to a local minimum on
the Born-Oppenheimer surface. The curvature energy is the energy
required to form a tubular form by folding the 2D triangular
network, and is expressed as $E_{c}(n,m)=
E_{b}(\textrm{2D})-E_{b}(n,m)$, in terms of the difference of the
binding energies in the closed-packed (111) atomic plane and in
the $(n,m)$ tube. For the  relaxed 2D triangular network we found
$E_{b}(\textrm{2D})=2.84$ eV/atom. Normally, $E_{c}(n,m)$
increases with decreasing $R$. The calculated curvature energies
comply with the expression obtained from classical elasticity
theory $E_{c}=Yw^{3}\Omega/24 R^{2}$ ($Y$: Young's modulus, $w$:
thickness of the tube, $\Omega$: atomic volume), and are fitted to
an expression $E_{c}=\alpha/R^{2}$ in Fig.~\ref{fig:f}.

All  the gold nanotubes given in Table~\ref{tab:res} are stable
when they are standing free. The cohesive energy values, $E_{b}$,
given in Table~\ref{tab:res}  gradually decreases with decreasing
$R$. The distribution of calculated Au-Au bond lengths in the
relaxed (5,5) tube has a sharp peak at $d=2.76$\AA,
  and relatively weaker individual peaks in the range of
$2.85<d<2.89$\AA. This distribution is, however, modified in the
(5,4) and (5,3) tubes, where the sharp peak at $d=2.76$\AA{} tends
to weaken and distribute in a wider range. The shortest $d$'s
correspond to the bonds forming the helical strands. As far as
applications as interconnect or nanodevice are concerned, it is
important to know whether the free-standing but finite length gold
SWNTs are stable. The structure optimization of free $4L$ long
(5,5) SWNT with open ends has resulted in a stable structure with
negligible rearrangements of atoms relative to the infinite tube.
Most importantly, the open ends have not been capped. The
stability of gold SWNTs against clustering is somewhat nontrivial,
though not totaly unexpected. Noting that free-standing finite
zigzag chains of gold are found to be stable \cite{sanchez}, the
present tubular structures are expected to be even more resistant
to clustering owing to their higher atomic coordination.

Among three SWNTs  with $n=5$, the $(5,5)$ tube is energetically
the most favorable. This situation is seemingly in disagreement
with the experiment indicating that $(5,3)$ gold tube is the
structure observed during the thinning process of gold nanowires
\cite{oshima03}. Nevertheless, this apparent contradiction is
reconciled by the fact that the calculations are performed for
free-standing infinite tubes, whereas the experiment is for a
finite tube stretching between two gold electrodes. Hence, $E_{b}$
should not be taken as a criterion to decide on the long-lived
metastable states of suspended nanowires.

Introducing the criterion of minimum string tension rather than
the total free energy for the stability of nanowires, Tosatti
\emph{et al.} \cite{tosatti01} have theoretically investigated a
class of gold nanowire structures having a single helical shell
covering a central linear strand of atoms. They found that the
wire having $(7,3)$ outer shell exhibits the minimum string
tension and was specified as ``magic", in agreement with
observation. Here, we carry out string tension analysis  for
single-wall gold nanotubes without a central strand. The string
tension $f$ of a nanowire is defined through the consideration of
the positive work done in drawing the wire out of the tips, and is
given by \cite{tosatti01}, $f=F-\mu N/L$.
Here, $F$ is free energy of the one unit cell of the wire. At zero
temperature, $F$ equals to the total energy $E$ of the wire; $\mu$
is the chemical potential of bulk gold, and is calculated to be
$\mu\simeq -3.2$ eV within GGA, in consistence with the
calculations made for the tubes. Calculated values of the string
tension do not exhibit a monotonic decrease as a function of $R$
as displayed in Fig.~\ref{fig:f}. In the plot one immediately
recognizes that $(5,3)$ and $(4,3)$ SWNTs have lower string
tension values as compared to those of their immediate neighboring
structures, thus they are favorable magic structures of gold
SWNTs. Aside from the reported (5,3) tube \cite{oshima03}, our
analysis predicts that the $(4,3)$ tube with $R=1.85$ \AA~is
another candidate for being a ``magic" structure which is not
observed yet. It appears that the (5,3) gold SWNT is favored,
since it lowers the tension exerted by two gold tips. In
principle, the string tension calculation and the geometric
relaxation of the structures should be performed
self-consistently. The $f$ values reported in Table \ref{tab:res},
however, are results of the first iteration obtained by using the
$F$ and $L$ values of the bare unstrained tubes. Nevertheless, we
tested that the reduction in $f$ after the full self-consistent
calculation is less than 0.3\% for the $(5,3)$ tube, too minor to
have any implications on our conclusions.

\begin{figure}
\includegraphics[scale=0.4]{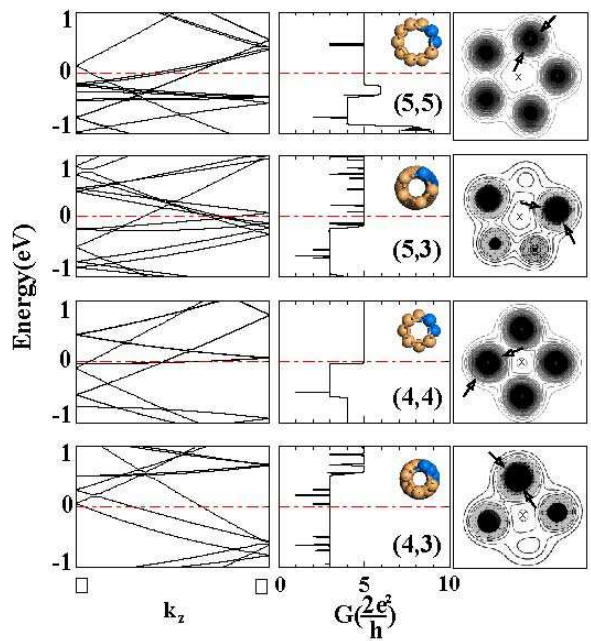}
\caption{Electronic energy-band structure and ballistic
conductance of infinite $(5,5)$, $(5,3)$, $(4,4)$, and $(4,3)$
gold SWNTs. The Fermi level is set to zero in each panel. Right
panels are contour plots of the total charge density on a plane
perpendicular to the tube axis. The charge density is negligible
at the centers of the tubes.} \label{fig:cond}
\end{figure}

\begin{figure}
\includegraphics[scale=0.4]{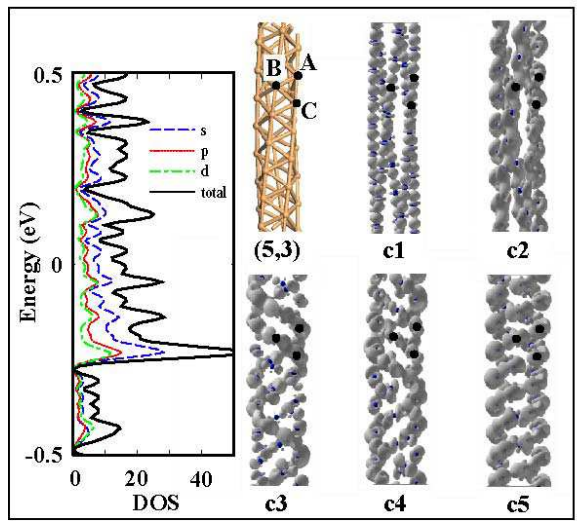}
\caption{Left: Total and orbital-decomposed densities of states
around the Fermi level ($E=0$) of $(5,3)$ gold SWNT. Right: A
schematic description of the structure, and isosurfaces for the
charge densities of the five states at the Fermi level
corresponding to current transporting channels c1, c2,\ldots c5.
Three reference points A, B, C are shown by dark spots on each
plot.} \label{fig:chan}
\end{figure}

Calculated energy band structures and  \textit{ab-initio}
 ballistic conductance plots of some infinite SWNTs  presented
in Fig.~\ref{fig:cond} are of particular interest. The tubular
character is demonstrated by contour plots of charge density,
$\rho_{T}({\rm \bf r})=\sum^{occ}_{i,k}\Psi^{*}_{i}({\rm \bf
r},k)\Psi_{i}({\rm \bf r},k)$.  Here $\rho_{T}({\rm \bf r})$ is
dramatically  different from that of gold nanowires with a strand
at the center. Bands near the Fermi level are derived mainly from
the Au-$6s$ orbitals; one band displays significant $5d_{z^2}$
hybridization. Flat $5d$-bands occur $\sim$1 eV below the Fermi
level. Despite 1D character of SWNTs, the bands which cross the
Fermi level do not allow for Peierls distortion. The character of
 states near the Fermi level of (5,3) tube is revealed by the density of
 states plots
in Fig.~\ref{fig:chan}. A single chain of gold atoms has unit
quantum conductance ({\it i.e.} $G_{0}=2e^{2}/h$)
\cite{ohnishi98,yanson98}. It has been usually contemplated that
one conductance channel is associated with each helical strand.
Accordingly, the ballistic conductance of $n$-strand gold nanotube
would be about $nG_0$. The three infinite SWNTs with $n=5$ have
indeed equilibrium conductance values of $5G_0$. However, of the
4-strand nanotubes $(4,4)$ also  has $5G_0$ conductance, while the
$(4,3)$ structure has only three channels for the ballistic
conductance. The 3-strand family of nanotubes have $3G_0$
conductance. Since the number of bands crossing the Fermi level
determines the conductance  of an infinite SWNT, there is no
direct correlation between numbers of strands and current
transporting channels, rather the cross section of the tube is
expected to be crucial in determining the number of channels.
Indeed, a minor reduction in the cross section area of the $(4,4)$
tube due to axial stretching reduces the number of channels from 5
to 3. Dips of size $G_0$ or $2G_0$ in the conductance plots are
another interesting feature one notes. They are due to small gap
openings in the energy band diagrams of these helical structures.
A recent calculation on the conductance of helical nanowires
attributes such characteristic dips to the non-circular
cross-section of the wires\cite{okamoto}. Note that the
conductance plots of (5,3) and (4,3) tubes in Fig.~\ref{fig:cond}
have more of those dips since they are more chiral. Finally, we
calculated the conductance of finite size (one unit cell) (5,3)
tube which is connected to two fcc gold electrodes through single
Au atoms from both ends, and found $G=1.75G_0$. Dramatic reduction
from $ 5G_0$ (the conductance of infinite tube)  is attributed to
the contacts with electrodes.

Prediction of chiral currents in chiral (also mechanically
stretched) single-wall carbon nanotubes has been of interest
because of the self-inductance and nanocoil effects
\cite{miyamoto}. In principle, chiral current passing through a
one-micron $(5,3)$ SWNT can induce magnetic fields of several
Tesla \cite{bagci}. According to Altshuler-Aronov-Spivak effect
\cite{altshuler} electrons circling a cylindrical conductor that
encloses a magnetic flux $\Phi$ give rise to a periodic
oscillating resistance as a function of $\Phi$. The observed
short-period oscillations in the magnetoresistance of multi-wall
carbon nanotubes have been attributed to chiral currents developed
by mechanical stretching of them \cite{bachtold}. It is of
interest to reveal whether the electrical current passing through
a $(5,3)$ SWNT has chirality due to the helical structure of the
tube. In Fig.~\ref{fig:chan}, charge density plots of five states
at the Fermi level, corresponding to the conduction channels,
clearly indicates the chirality. The geometry of the atomic
positions in the $(5,3)$ tube provides three distinct
circumferential directions for the flow of the current. These
directions can be defined as AB, AC, and BC in terms of the
reference points A, B, C depicted in Fig.~\ref{fig:chan}. While
the motion along  AB and AC directions corresponds to right-handed
helices, the helical path along the BC direction is left-handed.
The period and direction of chirality are different for different
channels; c3 and c4 have opposite chiral directions compared to
c1, c2, and c5. Hence, the resultant chirality effect may be
weaker than one expects. For the infinite tube, being a
transmission eigenchannel of the system, each channel contributes
a unit quantum conductance. However, for suspended tubes, the
chirality of the net current depends on the combined ``nanocoil"
effects of the conduction channels, as well as on the contacts.

In conclusion, we showed that free-standing gold chiral $(n,m)$
tubes with $3\leq n \leq 5$  are stable and exhibit novel
electronic and transport properties. Our analysis explains why the
experimentally observed (5,3) tube suspended between two gold tips
is favored, and indicates that the string-tension criterion
introduced by Tosatti \textit{et al} \cite{tosatti01} is also
valid for tubular structures. Using this criterion we predict that
tip-suspended (4,3) chiral gold tube is another structure that can
 be observed. We found that there is no direct correlation
between the numbers of conduction channels and helical strands
making the tubular structure. Current transporting states display
different periods and chirality, the combined effects of which
lead to weaker chiral currents on SWNTs.


\begin{references}
\bibitem{gimzewski} J. K. Gimzewski and R. M\"{o}ller, Phys. Rev. B \textbf{36}, 1284 (1987).
\bibitem{ciraci89} S. Ciraci and E. Tekman, Phys. Rev B \textbf{ 40}, 11969 (1989).
\bibitem{agrait93} N. Agra\"{\i}t \textit{et al.}, Phys. Rev. B {\bf 47}, 12345 (1993); N. Agra\"{\i}t \textit{et al.}, Phys. Rev. Lett. \textbf{74}, 3995 (1995).
\bibitem{pascual93} J. I. Pascual \emph{et al.},  Phys. Rev. Lett. {\bf 71} 1852 (1993).
\bibitem{mehrez97} H. Mehrez and S. Ciraci, Phys. Rev. B {\bf 56}, 12632 (1997).
\bibitem{gulseren98} O. G\"{u}lseren  \emph{et al.}, Phys. Rev. Lett. {\bf 80}, 3775 (1998).
\bibitem{sorensen98} M. R. Sorensen  \emph{et al.},  Phys. Rev. B {\bf 57}, 3283 (1998).
\bibitem{kondo00} Y. Kondo and K. Takayanagi, Science {\bf 289}, 606 (2000).
\bibitem{oshima03} Y. Oshima  \emph{et al.},  Phys. Rev. Lett. {\bf 91}, 205503 (2003).
\bibitem{tosatti01} E. Tosatti \emph{et al.}, Science {\bf 291}, 288 (2001).
\bibitem{ohnishi98} H. Ohnishi  \emph{et al.},  Nature (London) \textbf{395}, 780 (1998).
\bibitem{yanson98} A. I. Yanson \emph{et al.}, Nature (London) \textbf{395}, 783 (1998).
\bibitem{oshima02} Y. Oshima \emph{et al.}, Phys. Rev. B \textbf{65}, 121401 (2002).
\bibitem{vasp} Calculations have been performed by using the
{\sc VASP} software: G. Kresse and J. Hafner, Phys. Rev. B
\textbf{47}, 558 (1993); G. Kresse and J. Furthm\"{u}ller,
\textit{ibid} \textbf{54}, 11169 (1996).
\bibitem{trans} The methodology of the {\sc transiesta-c} software  is based on: M. Brandbyge \emph{et al.}, Phys. Rev. B \textbf{65}, 165401 (2002). The software is provided by Atomistix Corp.
\bibitem{sanchez} D. S{\'a}nchez-Portal \emph{et al.}, Phys. Rev.
Lett. \textbf{83}, 3884 (1999).
\bibitem{okamoto}M. Okamoto \textit{et al.}, Phys. Rev. B \textbf{64}, 033303 (2001).
\bibitem{miyamoto} Y. Miyamoto \textit{et al.}, Phys. Rev. B \textbf{60}, 13885 (1999).
\bibitem{bagci} V. M. K. Bagci \emph{et al.}, Phys. Rev. B \textbf{66}, 045409 (2002).
\bibitem{altshuler} B. L. Altshuler \emph{et al.}, JETP Lett. \textbf{ 33}, 94 (1981).
\bibitem{bachtold} A. Bachtold {\it et al.}, Nature (London) \textbf{397}, 673 (1999).

\end{references}
\end{document}